\documentclass{ws-procs9x6}
\usepackage{epsfig}
\input epsf
\def\Journal#1#2#3#4{{#1} {\bf #2}, #3 (#4)}

\def\EPJ{{\em Eur. Phys. J.} {\bf C}}
\def\PREP{\em Phys. Rept.}

\def\RNC{\em Riv. Nuovo Cim.}

\newcommand{\gevcc}{\ensuremath{{\rm GeV}\!/c^2}}

\newcommand{\epemto}{\ensuremath{{{\rm e}^+{\rm e}^- \to}}}

\newcommand{\st}{\ensuremath{{\tilde{\rm t}}}}

\newcommand{\sq}{\ensuremath{{\tilde{\rm q}}}}
\newcommand{\glu}{\ensuremath{{\tilde{\rm g}}}}
\newcommand{\mst}{\ensuremath{m_\st}}

\newcommand{\mglu}{\ensuremath{m_\glu}}

\newcommand{\sqsqbar}{\ensuremath{\sq\bar{\sq}}}
\newcommand{\qqbar}{\ensuremath{{\rm q\bar{q}}}}

\def\be{\begin{equation}}
\def\ee{\end{equation}}
\def\bea{\begin{eqnarray}}
\def\eea{\end{eqnarray}}

\begin{document}

\title{SUSY searches at LEP}

\author{ A.C. Kraan}

\address{University of Pennsylvania, Department of Physics and Astronomy, \\ 209 S 33rd Street, PA 19104, Philadelphia, USA}

\maketitle\abstracts{
Searches for SUSY particles have been performed in 3.6 fb$^{-1}$ e$^+$e$^-$ data collected by the LEP detectors at $\sqrt{s}$ between 90 GeV and 209 GeV. This talk reviews some of the relevant searches for SUSY particles at LEP. No excess of events is observed in any channel. Results are interpreted in the context of the MSSM.}

\section{Introduction}
Supersymmetry~\cite{susy} (SUSY) is an attractive candidate as theory for physics beyond the Standard Model (SM). In the simplest SUSY implementation, the  Minimal SuperSymmetric Model (MSSM), containing the minimal number of additional particles, each SM particle has a supersymmetric 'partner', differing in spin by $1/2$. The superpartners of fermions, gauge bosons and the two MSSM Higgs doublets are called \emph{sfermions}, \emph{gauginos} and \emph{Higgsinos}, respectively. 

An important quantity in SUSY phenomenology is R-parity. In R-parity conserving (RPC) models, sparticles are only pair produced, and the Lightest SUSY particle (LSP) is stable. Most naturally the LSP is a \emph{neutralino} $\tilde{\chi}_0$ (a mixture of the neutral gauginos and higgsinos), or sneutrino $\tilde{\nu}$ (superpartner of neutrino), although the latter is cosmologically disfavoured.

Widely accepted frameworks are the constrained Minimal SuperSymmetric Model (cMSSM) and the Minimal supergravity (mSUGRA). In the former masses and couplings can be derived from a few parameters: $\tan\beta$, the ratio of the vacuum expectation value of the two Higgs doublets; $\mu$, the Higgs mass parameter, $M_2$, the EW scale common gaugino mass; $m_0$, the GUT scale common scalar mass; and $A_0$, the trilinear couplings which enter in the prediction of the sfermion mixing. In the even more constrained mSUGRA framework the parameters are: $\tan\beta$; the sign of $\mu$; $m_0$; $m_{1/2}$, the GUT scale common gaugino mass that replaces $m_2$; and $A_0$, the GUT scale common trilinear coupling.

To search for SUSY, LEP data have been analysed. Results shown here are mainly based on the second phase of running, LEP\,2, when roughly 775 pb$^{-1}$ of data per experiment was collected at $\sqrt{s}$ of 130-209 GeV. 

This paper reviews a selection of LEP SUSY searches. First searches for sleptons, squarks and charginos  are summarized, and a limit on the neutralino LSP is shown. When available the SUSY LEP working group~\cite{susywg} results are used, based on combinations of ALEPH, DELPHI, L3 and OPAL (ADLO).  No signal has been observed and results will be given in the form of 95\% C.L. exclusion domains in the space of the relevant parameters. Also, DELPHI and ALEPH searches for a gluino LSP will be reviewed. Finally a few remarks will be given about alternative SUSY models.

\section{Slepton searches}
\begin{figure}[t]
\vspace{-0.2cm}
\centerline{\epsfxsize=1.7in\epsfbox{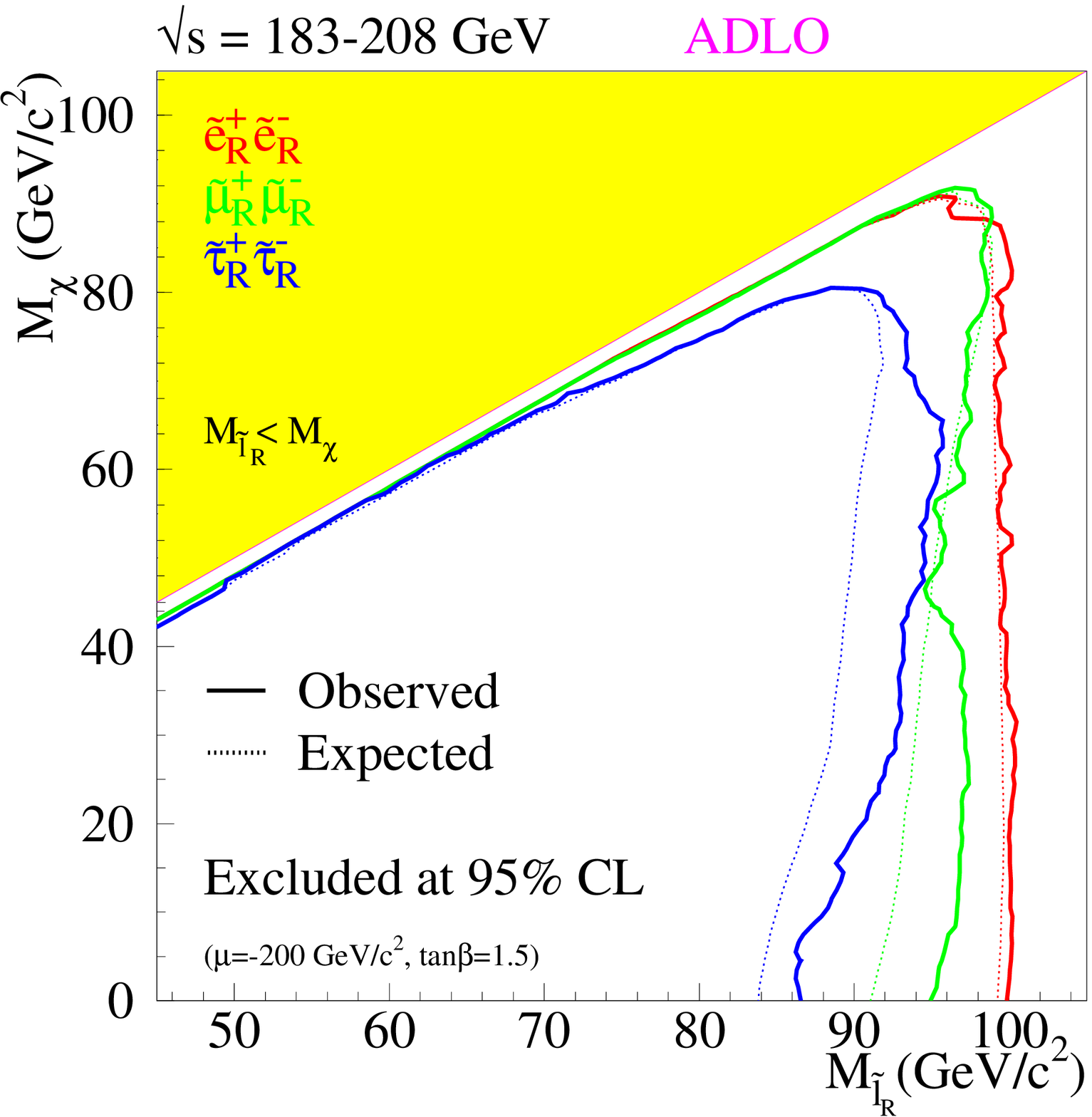}\hspace*{-0.2cm}\raisebox{-0.2cm}{\epsfxsize=1.7in\epsfbox{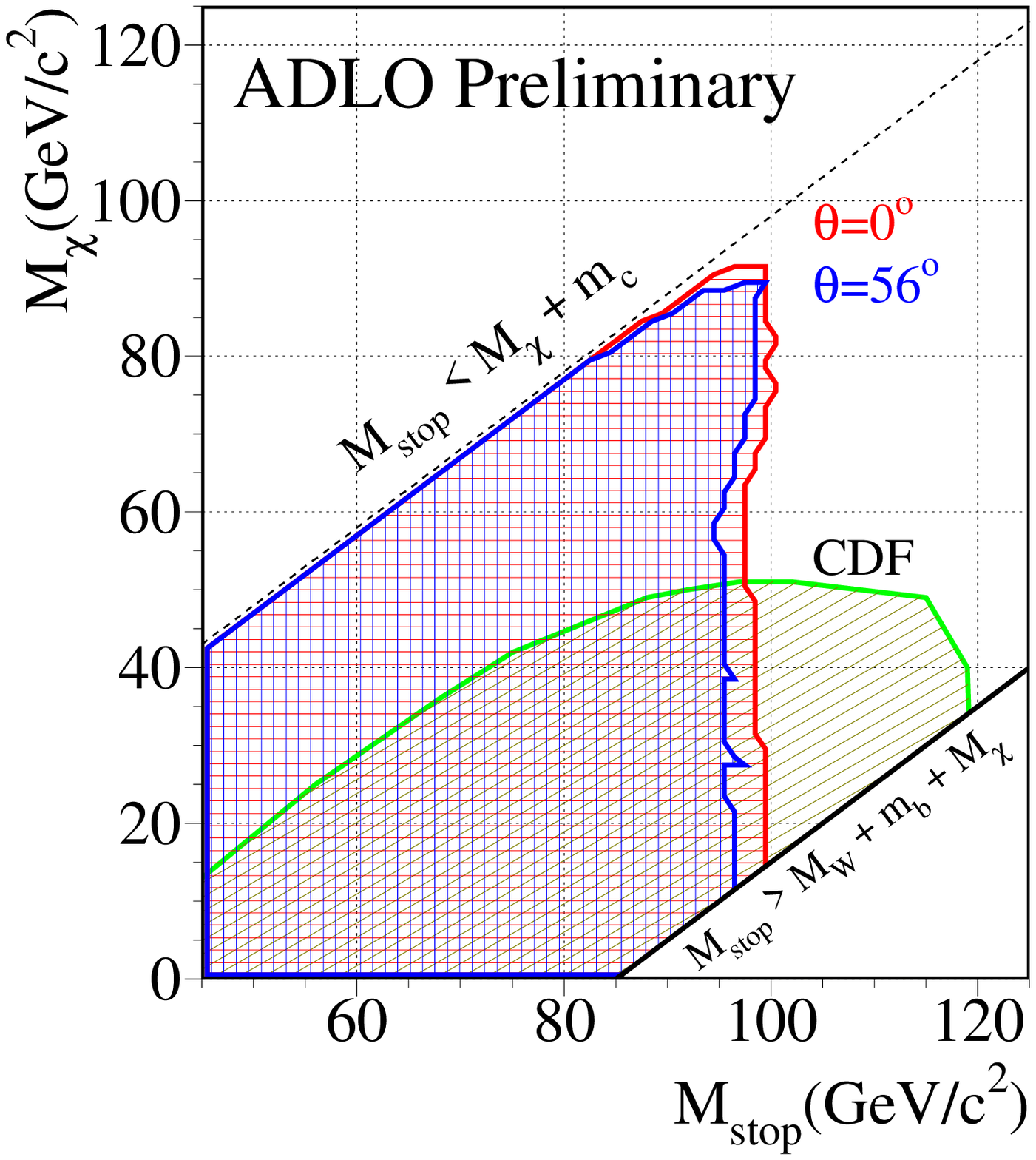}\hspace*{-0.8cm}\epsfxsize=1.7in\epsfbox{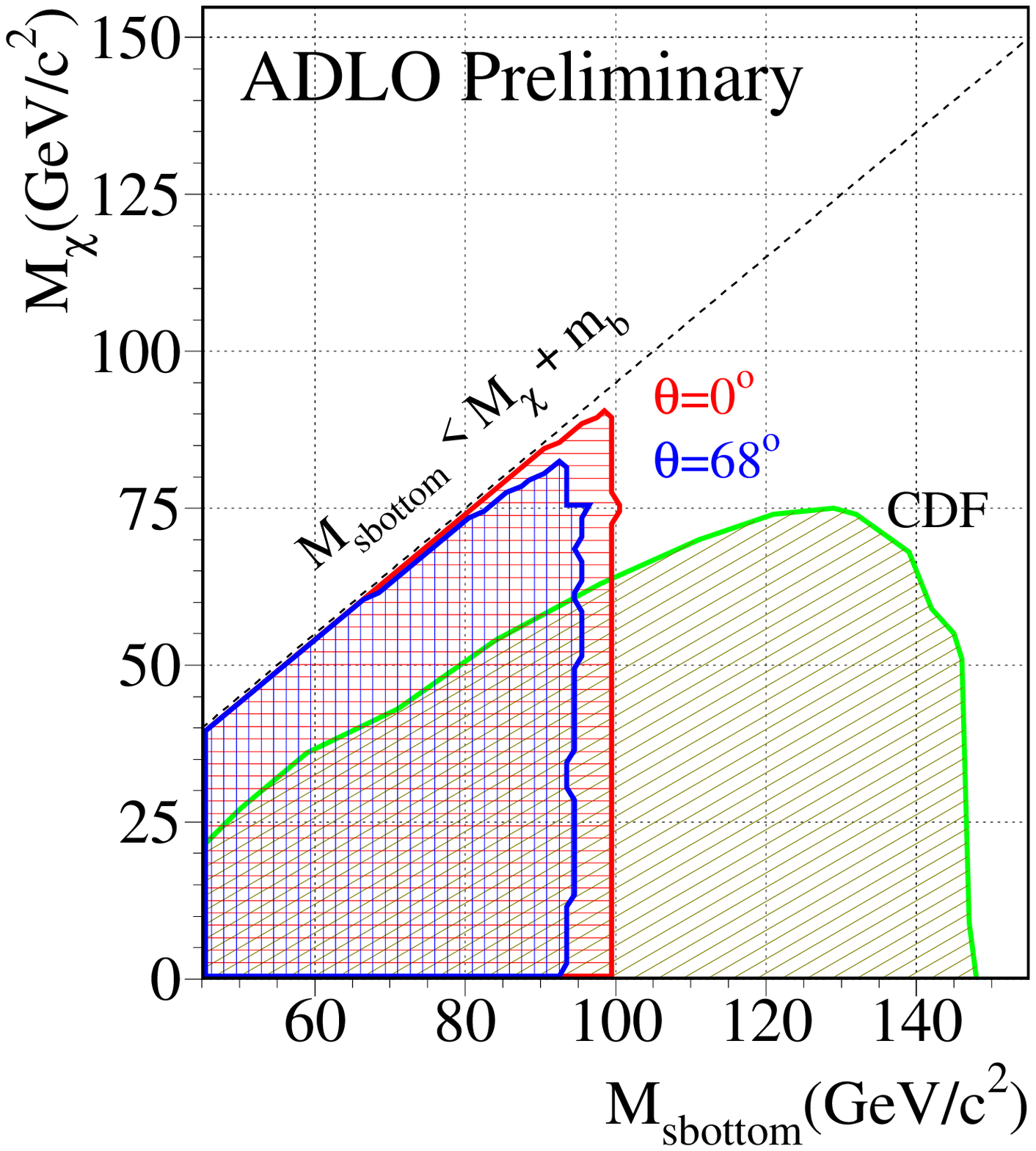}}    }
\caption{Left: excluded region (final) for pair produced sleptons decaying via $\tilde{l}\rightarrow l+\chi_0$. The parameter space point is $\mu = -200$ GeV$/c^2$, $\tan(\beta) = 1.5$. Middle: excluded regions  (preliminary) for pair produced stops decaying via $\tilde{t}\rightarrow c \chi_O$. Right: excluded regions (preliminary) for pair produced sbottoms decaying via $\tilde{b}\rightarrow b \chi_O$. Only squark masses above 50 \gevcc\ are shown, since lower values have been excluded at LEP\,1. \label{fig:sleptons}}
\end{figure}

If $m_0$ is small, the sleptons ($\tilde{l}$, SUSY partners of the leptons) can be light. In particular the \emph{stau} ($\tilde{\tau}$, SUSY partner of the $\tau$) may be light, because for the 3rd generation sfermions large mixing can occur between ``left-'' and ``right-handed'' partners, resulting in a heavy and a light mass eigenstate, the latter possibly in the reach of LEP. Sfermions would be pair produced via s-channel Z/$\gamma$ exchange, whereby for \emph{selectrons} ($\tilde{e}$, superpartners of electrons) t-channel $\tilde{\nu}$ exchange contributes also. The four experiments have searched for sleptons decaying into a lepton and a $\tilde{\chi}_0$ LSP, where the signature was two acoplanar leptons and missing energy due to the two escaping LSP's. The missing energy signal is closely related to the mass difference $\Delta$M between the slepton and the LSP: small $\Delta$M values imply large missing energy and vice versa. No excess of events was observed, and in Fig.~\ref{fig:sleptons} the mass limits are shown.

\section{Squark searches}
Due to sfermion mixing, the third generation squarks may be light. Squarks would be produced via s-channel Z/$\gamma$ exchange. Decay channels with a $\tilde{\chi}_0$ and $\tilde{\nu}$ LSP have been studied. In the channels $\tilde{t}\rightarrow c \tilde{\chi}^o$ and $\tilde{b}\rightarrow b + \tilde{\chi}^o$, the topology is missing energy and acoplanar quark jets. In the channel $\tilde{t}\rightarrow b\l\tilde{\nu}$, isolated leptons are additionally searched for. The four-body decay channel $\tilde{t}\rightarrow b\bar{f}_1f_2\tilde{\chi}^o$, characterised by multi-jets and missing energy, was studied by ALEPH. No excess of events was observed in any channel, and Fig.~\ref{fig:sleptons} shows examples of excluded regions.  The case in which the squark has a sizable lifetime (small $\Delta$M), is addressed by ALEPH, and resulted in an absolute limit on the stop mass of 63 \gevcc\, for a $\tilde{\nu}$ or $\tilde{\chi}_0$ LSP.

\section{Chargino searches}
The charged gauginos and Higgsinos mix to form \emph{charginos}, $\tilde{\chi}^{\pm}$, which would be pair-produced via s-channel Z/$\gamma$ exchange and t-channel $\tilde{e}$-exchange. If $m_0$ is small the sleptons are light, and thus the decay $\tilde{\chi}^{\pm}\rightarrow\tilde{l}\nu\rightarrow\tilde{\chi}_0$ could dominate. Also, light sleptons can lead to t-channel destructive interference in the production cross section. At large $m_0$, the search channel is $\tilde{\chi}^{\pm}\rightarrow\tilde{\chi}^0W^{\pm}$ only, and in that case the topology, besides missing energy, depends on the decay mode of the $W$. Searches have been done for different values of $\Delta$M between $\tilde{\chi}^{\pm}$ and $\tilde{\chi}^0$: prompt decays, long-lived and stable charginos have \begin{figure}[t]
\begin{center}
\epsfig{file=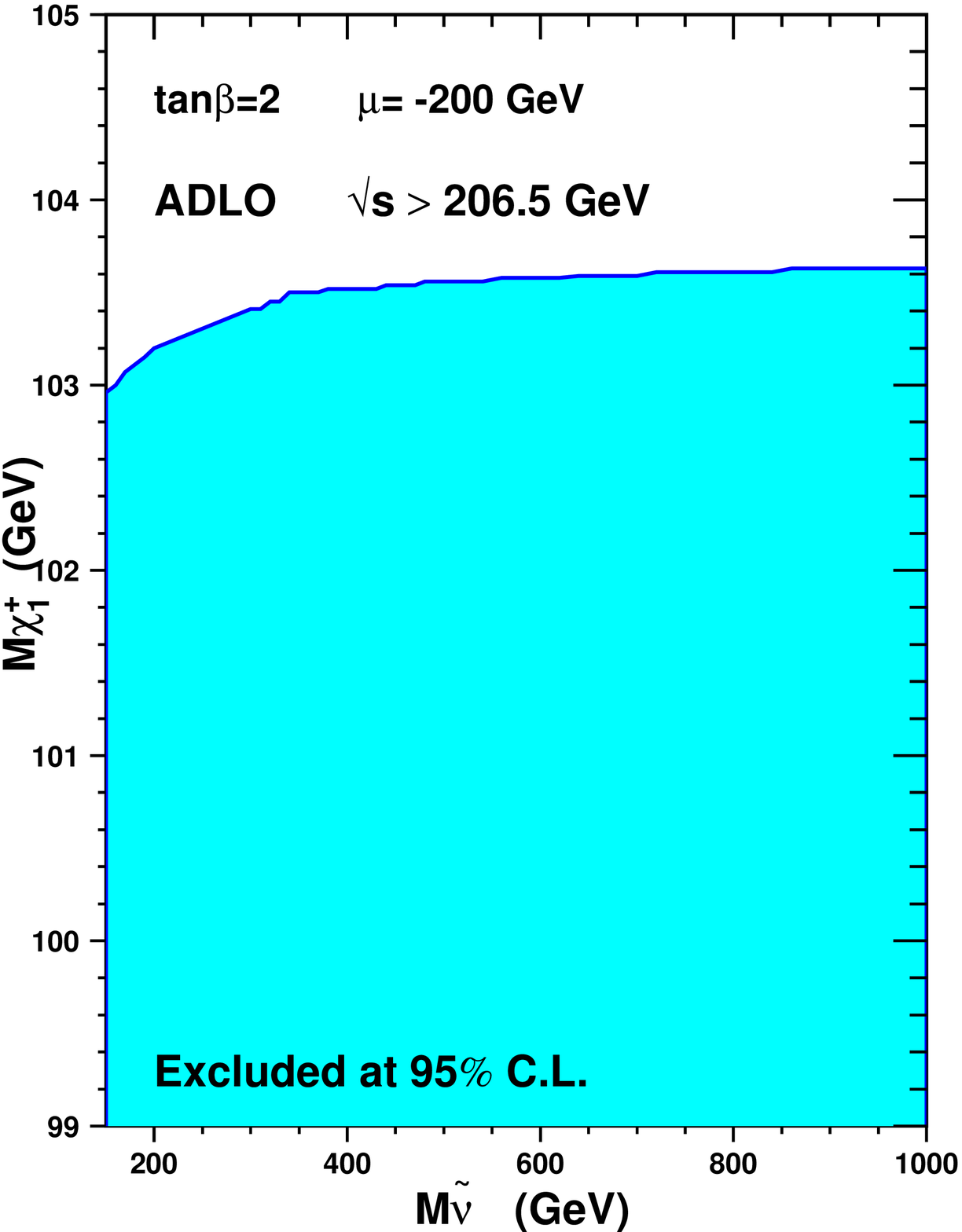,height=4cm,width=3.8cm}\hspace{-0.25cm}\raisebox{-0.2cm}{\epsfig{file=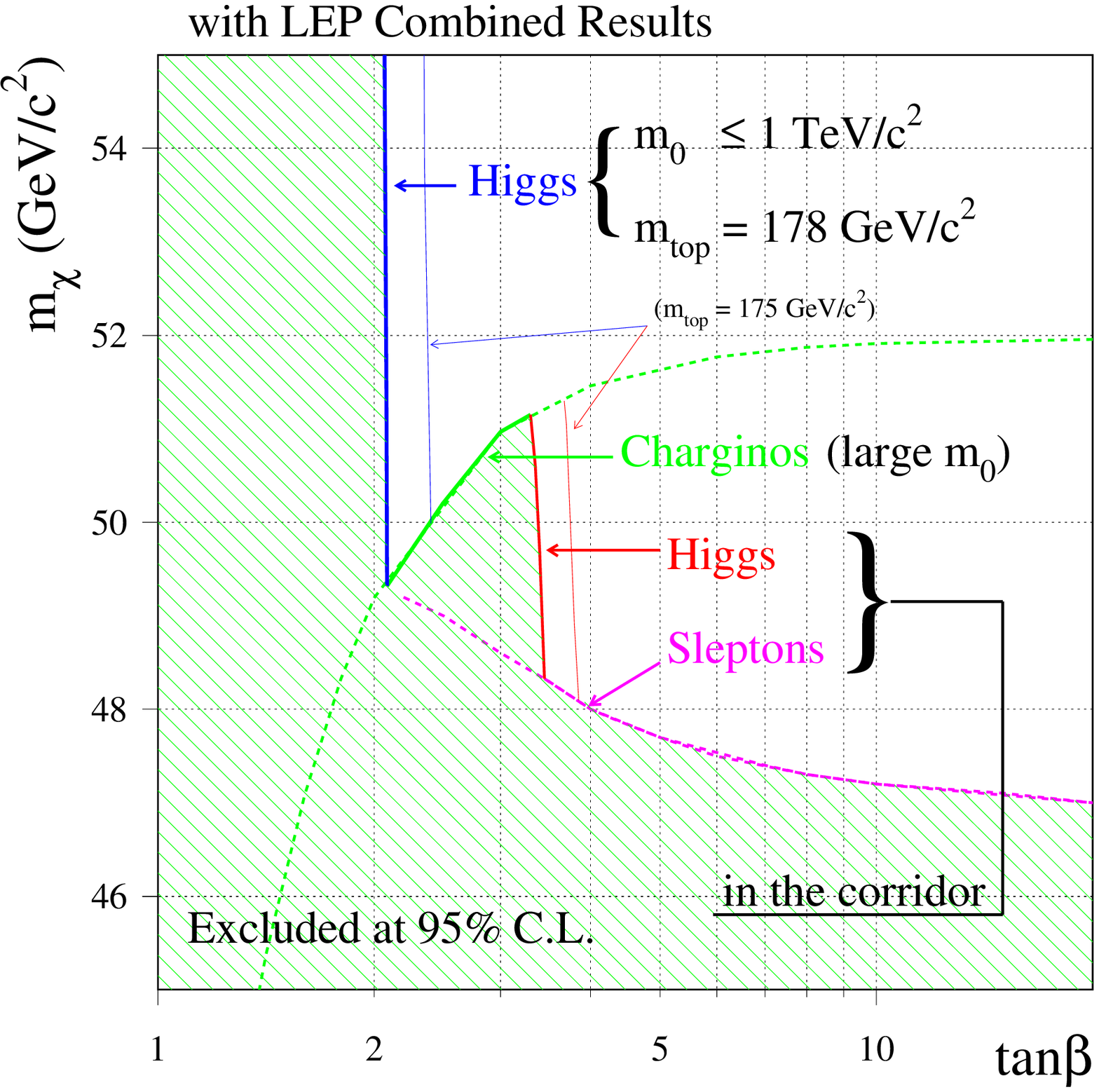,height=4.3cm,width=4.1cm}}\raisebox{-0.2cm}{\hspace{-0.25cm}\epsfig{file=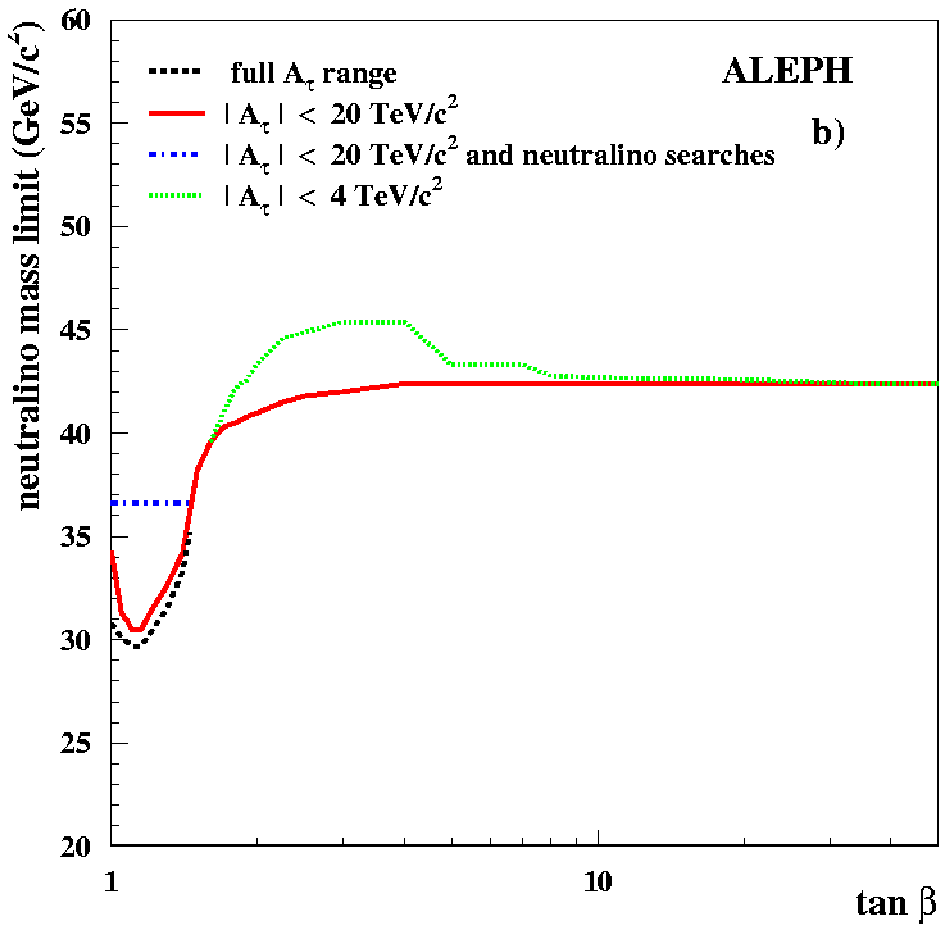,height=4.2cm,width=3.8cm}}
\end{center}
\caption{\footnotesize Left: exclusion plot of charginos as function of sneutrino mass (preliminary result). Middle: mass limit on the neutralino LSP as function of $\tan\beta$ (final result). Right: mass limit on the neutralino LSP as function of $\tan\beta$ including $\tilde{\tau}$-mixing (final).\label{fig:charginoss}}
\end{figure}been searched for. In Fig.~\ref{fig:charginoss} the exclusion plot is shown. 
\section{LSP limits}
The above searches can be combined with the cMSSM Higgs boson searches, and Fig.~\ref{fig:charginoss} shows results. In the cMSSM framework the lower mass limit on the LSP is 47 GeV. For mSUGRA (not shown) the limit is 50 GeV. The effect of $\tilde{\tau}$ mixing in the cMSSM, possibly causing the stau mass to be in between the $\tilde{\chi}^0$ and the $\tilde{\chi}^{\pm}$, has been investigated by ALEPH. Searches have been done for $e^+e^-\rightarrow\tilde{\chi}^+\tilde{\chi}^-\rightarrow \tilde{\tau}\nu\tilde{\tau}\nu\rightarrow\tau\nu\tilde{\chi}^0\tau\nu\tilde{\chi}^0$, $e^+e^-\rightarrow\tilde{\chi}^0_2\tilde{\chi}^0_1\rightarrow \tilde{\tau}\tau\tilde{\chi}^0_1\rightarrow\tau\tilde{\chi}^0_1\tau\tilde{\chi}^0_1$ and  $e^+e^-\rightarrow\tilde{\chi}^0_2\tilde{\chi}^0_2\rightarrow \tilde{\tau}\tau\tilde{\tau}\tau\rightarrow\tau\tau\tilde{\chi}^0_1\tau\tau\tilde{\chi}^0_1$, where in all cases missing energy and a number of taus was searched for. Fig.~\ref{fig:charginoss} shows the LSP limit as function of $\tan\beta$ when $\tilde{\tau}$ mixing is included. 

\section{Searches for gluino LSP} 
\begin{figure}[t]
\begin{center}
\vspace{-0.2cm}
\epsfig{file=peq05.epsi,height=3.4cm,width=3.4cm} \raisebox{-0.3cm}{\epsfig{file=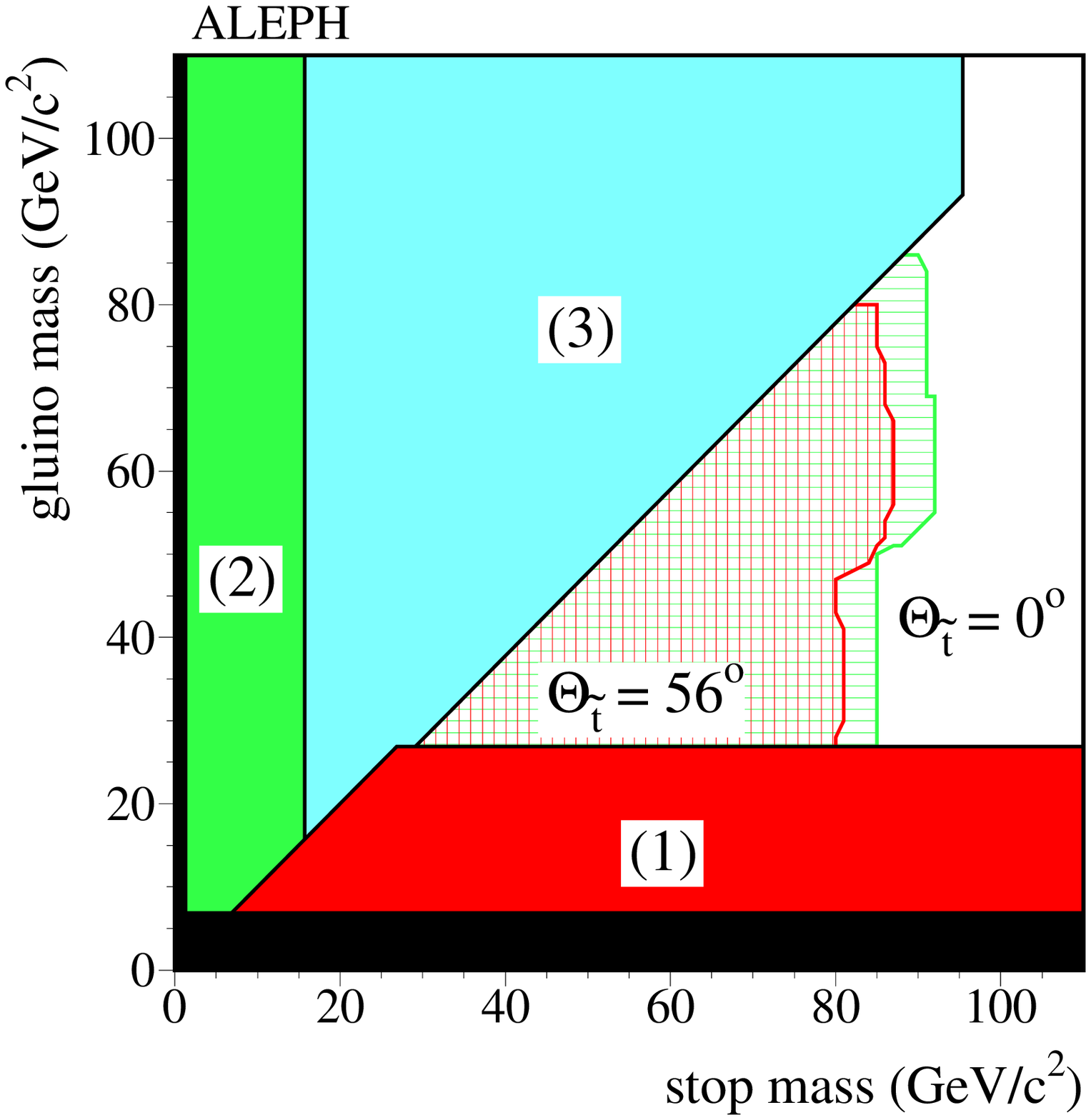,height=4.4cm,width=4.4cm}}\epsfig{file=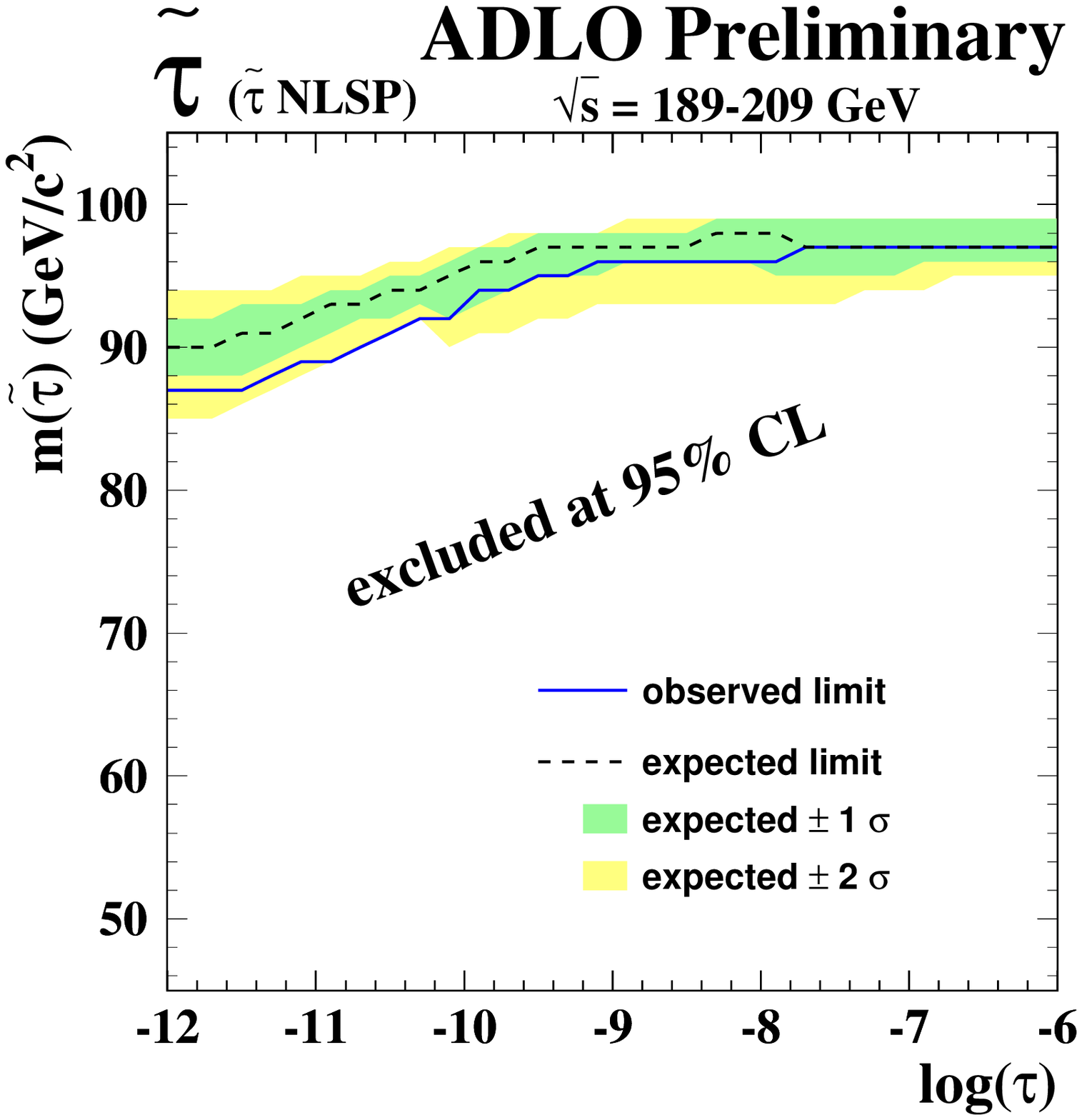,height=3.8cm,width=3.5cm}
\end{center}
\vspace{-0.2cm}
\caption{\footnotesize  Left: excluded area for $\tilde{b}\rightarrow b + \tilde{\chi}^o$ (DELPHI). Middle: excluded regions in the plane 
(\mst, \mglu) by ALEPH. Black area: excluded by the Z-lineshape. Regions (1), (2) and (3): excluded by the LEP\,1 search for \epemto\ \qqbar\glu\glu, the LEP\,1 search for  for \epemto\ \qqbar\sqsqbar, and the LEP\,2 search for stable squarks, respectively. Hatched area: excluded by the acoplanar jet plus missing energy search at LEP\,2. Right: Excluded mass regions for the $\tilde{\tau}$ mass as function of its lifetime for GMSB scenarios.\label{fig:all}}
\end{figure}

In contrast to conventional SUSY models predicting heavy gluinos decaying promptly, models exist with a \emph{gluino} ($\tilde{g}$, superpartner of the gluon) LSP. A stable gluino would hadronize into stable charged and neutral particles called R-hadrons, interacting hadronically and possibly electromagnetically in the detector.
Searches for gluino LSP's have been done by DELPHI and ALEPH. Gluinos would be produced via squarks, $e^+e^-\rightarrow\tilde{q}\tilde{q}$. The decay channels $\tilde{t}\rightarrow c + \tilde{g}$ and  $\tilde{b}\rightarrow b + \tilde{g}$ have been studied. The signature is two acoplanar jets and missing energy (due to the missing mass of the particle plus the poor hadronic interaction in the detector~\cite{aafke1}). 
Figure 1 displays the results for the decay channels $\tilde{b}\rightarrow b + \tilde{g}$, studied only by DELPHI~\cite{delphi}. The channel $\tilde{t}\rightarrow c + \tilde{g}$ has been studied by DELPHI and ALEPH. The ALEPH search is based on a combination of four different analyses: 1) the Z hadronic width, resulting in an exclusion of gluino masses below 6.3\,\gevcc and squark masses below 1.3\,\gevcc; 2) LEP\,1 searches for acoplanar jets and missing energy; 3) stable squark search with LEP2 data; 4) a direct search for $\tilde{t}\rightarrow c + \tilde{g}$, where the search topology is again missing energy and acoplanar jets. Issues addressed in the Monte Carlo simulation are stop hadronization, stop decay inside the stop-hadron, gluino hadronization into R-hadrons and the R-hadronic interaction in the detector~\cite{aafke2}. The result of the interplay of the four analyses is shown in Fig.~\ref{fig:all}. The ALEPH analyses allowed a mass limit of 80 GeV to be set on the stop mass in case of a gluino LSP.
\section{Other SUSY searches}
In GMSB models, the LSP is naturally the gravitino, while the neutralino or the slepton is the next-to-LSP (NSLP). Different NLSP decay lengths have been studied. Fig.~\ref{fig:all} shows the excluded mass regions in the case of a $\tilde{\tau}$ NLSP.
Searches for RPV topologies have also been done at LEP. Contrary to RPC models, the phenomenology of R-parity violating (RPV) models is different, because the LSP decays into SM particles.  Since LSP decaying into SM particles, the usual missing energy signal is absent, and instead, multi-jets and multi-leptons are searches for. No excess of events was observed.

\section*{Acknowledgements}
I thank Giacomo Sguazzoni for many useful comments for the talk and for this paper. Thanks also to Stefania, who fortunately could not come to the conference, and to Paolo Azzurri for nicely reviewing this paper.

\end{document}